\documentclass[preprint,12pt,authoryear]{elsarticle}

\usepackage{amsmath,bbm,hyperref,mathrsfs,graphicx}

\journal{Physics Letters B}

\begin{document}

\begin{frontmatter}

\title{$\mathcal{PT}$ -Symmetric Quantum Field Theory in Path Integral Formalism and Arbitrariness Problem}

\author[label1]{Yi-Da Li\fnref{fn}}
\fntext[fn]{liyd20@mails.tsinghua.edu.cn}
\author[label1,label2]{Qing Wang\corref{cor}}
\cortext[cor]{wangq@mail.tsinghua.edu.cn}
\affiliation[label1]{organization={Department of Physics, Tsinghua University},
            city={Beijing},
            postcode={100084}, 
            country={P. R. China}}
\affiliation[label2]{organization={Center for High Energy Physics, Tsinghua University},
            city={Beijing},
            postcode={100084}, 
            country={P. R. China}}

\begin{abstract}
Perturbative $\mathcal{PT}$-symmetric quantum field theories with anti-Hermitian and $\mathcal{P}$-odd interaction terms are studied in path integral formalism and the $i\phi^3$ model is calculated in detail. The nonlocal field transformation induced by the $\mathcal{C}$ operator and corresponding transformations to Hermitian theories are given systematically, which are  manifestly 4-dimensional invariant. It is found that the $i\phi^3$ model can be transformed into infinitely many physically inequivalent Hermitian theories with different $S$ matrices under above nonlocal transformations. Similar problem caused by nonlocal tranformations in Hermitian quantum field theories is also discussed. $0$-dimensional models are studied numerically to verify the validity of those transformations. We put forward the question of physical meaning of $\mathcal{PT}$-symmetric quantum field theory based on the arbitrariness caused by the seemingly inevitable nonlocality.
\end{abstract}

\begin{keyword}
$\mathcal{PT}$ symmetry \sep quantum field theory \sep path integral \sep arbitrariness problem
\end{keyword}

\end{frontmatter}

\section{Introduction\label{sec-intro}}

Roads towards new physics are broadened to a great extent by allowing Hamiltonians to be non-Hermitian but $\mathcal{PT}$-symmetric\cite{bender1998}, or equivalently, pseudo-Hermitian\cite{ali2002-1,ali2002-2,ali2002-3,ali2003}. $\mathcal{PT}$(parity and time) symmetry is discovered by studying the quantum-mechanical Hamiltonian\cite{bender1998} $H=p^2+m^2x^2-(ix)^N$. It is intriguing that this Hamiltonian has a real and bounded below spectrum\cite{dorey2001} if  defined properly. The $N=4$ case is especially important for particle physics because the effective potential of Higgs fields runs into the form of $\lambda(\phi)\phi^4$ with $\lambda(\phi)<0$ in high energy which renders the standard model vacuum unstable\cite{higgs}, and the stabilization of $-x^4$ potential in quantum mechanics provides a first-step solution to this problem without introducing extra fields\cite{bender2011}.

Real spectra of  $\mathcal{PT}$-symmetric Hamiltonians come with non-trivial inner product in Hilbert space. To be concrete, there exists an extra symmetry operator $\mathcal{C}$ for a $\mathcal{PT}$-symmetric Hamiltonian, and the time evolution is unitary only in the $\mathcal{CPT}$ inner product\cite{bender2002} instead of the usual Hermitian inner product. And a non-Hermitian $\mathcal{PT}$-symmetric Hamiltonian can be transformed into Hermitian forms making use of $\mathcal{C}$. However, presence of the $\mathcal{C}$ operator makes calculation of physical quantities difficult because the form of $\mathcal{C}$ is complicated in interacting theories\cite{bender2003}. Further more, $\mathcal{C}$ is generally nonlocal in $\mathcal{PT}$-symmetric quantum field theories\cite{bender2004a}. Nonlocality in quantum field theory causes various problems. It has been shown in \cite{novikov2019} that the 2 to 2 scattering amplitude in the $i\phi^3$ model, which is calculated in interaction picture, violates causality. In this paper, we show that $i\phi^3$ type\footnote{$i\phi^3$ model is a special case where the interaction part of Hamiltonian or Lagrangian is anti-Hermitian and odd in $\mathcal{P}$ while the free part is Hermitian and even in $\mathcal{P}$.} $\mathrm{PT}$-symmetric quantum field theories can be transformed into infinitely many physically inequivalent Hermitian theories with different $S$ matrices, which obscures their physical meanings.

Actually, there are studies about the $i\phi^3$ model decades ago, such as those in the study of Yang-Lee edge singularities\cite{yanglee}. After the discovery of $\mathcal{PT}$-symmetry\cite{bender1998}, the $i\phi^3$ model regains attentions and is investigated in many aspects including calculation of the nontrivial inner product\cite{bender2004a,bender2004b}, comparison with ordinary $\phi^3$ model\cite{bender2012b}, vacuum stability\cite{shalaby2017} and more\cite{bender2013,shalaby2019,shalaby2020,bender2005,aditya2021}. But if the $i\phi^3$ model is to serve as a model for some kind of particles which do not interact with external sources, the new inner product rather than the ordinary Dirac inner product is needed to guarantee unitarity. In this case, we must specify the equivalent Hermitian Hamiltonian for the $i\phi^3$ model to be able to calculate observables such as the $S$-matrix. \cite{novikov2019} did this work in the Hamiltonian framework, and violation of causality is concluded. Here we continue the issue and show that the uniqueness of equivalent Hermitian Hamiltonians for the $i\phi^3$ model is even problematic, as stated in the past paragraph.

To make formulae clean, we develop representations of $\mathcal{C}$ operator and corresponding transformations to Hermitian theories in path integral formalism. Calculating physical correlation functions in path integral formalism has been studied in \cite{jones2007}. It has been shown that the effect of $Q$ operator defined by $\mathcal{C}=\mathcal{P}e^{-2Q}$ in our language is embedded in the relaltions between fields in $\mathcal{PT}$-symmetric theories and that in the corresponding Hermitian ones, but the form of $Q$ used in \cite{jones2007} is still inherited form canonical formalism thus complicating calculations and burying manifest 4-dimensional invariance owing to the presence of generalized momentum. As far as we know, there is no such discussion without using generalized momentum which is absent in path integral defined by Lagrangian from the beginning.

This paper is organized as follows. In Sec. \ref{sec-pt} we briefly review $\mathcal{PT}$ symmetry in quantum theory. In Sec.\ref{sec-path}, we reformulate basic elements of $\mathcal{PT}$-symmetric quantum field theory such as $\mathcal{C}$ operator, as transformations in path integral formalism. In Sec. \ref{sec-infinite}, we show that there are infinite inequivalent Hermitian theories relating to $i\phi^3$ model by nonlocal transformations, which is our key result. In Sec. \ref{sec-herm}, similar problem arising in Hermitian $\lambda\phi^4$ model is analyzed. In Sec. \ref{sec-0}, we analyze the path integral of $0$-dimensional $i\phi^3$ and numerically show the existence of above transformations in their $0$-dimensional versions. In Sec. \ref{sec-final}, we conclude and discuss possible developments in future.

\section{$\mathcal{PT}$ Symmetry in Quantum Theory\label{sec-pt}}

For readers' convenience, we make a brief review of  $\mathcal{PT}$ symmetry. See more details in\cite{bender-review,bender-book}. A Hamiltonian $H$ is said to be $\mathcal{PT}$-symmetric if it commutes with $\mathcal{PT}$, where $\mathcal{P}$ is the  parity operator and $\mathcal{T}$ is the time reversal operator. A generalized $\mathcal{PT}$ symmetry can be an antilinear symmetry\cite{bender2002-gen}. It has been shown\cite{ali2002-3} that having an antilinear symmetry is equivalent to pseduo-Hermiticity which requires that $H$ is similiar to its Hermitian conjugation for some invertible operator $\rho$
\begin{equation}
\rho H\rho^{-1}=H^\dagger.
\end{equation}
If $H$ has an entirely real spectrum which is also to say the $\mathcal{PT}$ symmetry is unbroken\cite{ali2003}, there exists a $\mathcal{C}$ operator that commutes with both  $\mathcal{PT}$ and $H$. $\rho$ can also be written in a manifest Hermitian form $\rho=\eta^\dagger\eta$ to serve as a positive-definite inner product which is exactly the $\mathcal{CPT}$ inner product\cite{bender2002} under which the time evolution by $H$ is unitary, and we can further define a Hermitian Hamiltonian $h=\eta H\eta^{-1}$ corresponding to $H$. To calculate physical quantities such as $S$ matrix elements one must use either Hermitian $h$ with the usual Hermitian inner product or  $\mathcal{PT}$-symmetric $H$ with the $\rho$ inner product.

In perturbation theory, it is convenient to consider Hamiltonians in the form $H=H_0+H_1$ where $H_0$ is Hermitian and $\mathcal{P}$-even while $H_1$ is anti-Hermitian and $\mathcal{P}$-odd such that $\mathcal{P}H\mathcal{P}^{-1}=H^\dagger$. The $\mathcal{C}$ operator can then be written as $\mathcal{C}=\mathcal{P}e^{-2Q}$ and perturbation series for $Q$ can be obtained systematically\cite{bender2004b}. And $\eta$ can be chosen as $\eta=e^{-Q}$ to calculate the Hermitian Hamiltonian $h=\eta H\eta^{-1}$. However, $\eta$ can differ from $e^{-Q}$ by an unitary operator and $\mathcal{C}$ is also shown to be nonunique\cite{bender2012} thus adds nonuniqueness to $\eta$ even if $\eta=e^{-Q}$ is chosen. In quantum field theory, $Q$ is shown to be nonlocal, and any local expression has not been found yet. It is the nonlocality of $Q$ that causes the arbitrariness problem of $\mathcal{PT}$-symmetric quantum field theory as will be shown in Sec. \ref{sec-infinite}.

In this paper we study $\mathcal{PT}$-symmetric $i\phi^3$ model in detail and other models of the above type can be generalized in the same manner. For the non-Hermitian $\mathcal{PT}$-symmetric Hamiltonian
\begin{equation}
H=\int d^3x\left[\frac12\pi^2_{\vec{x}}+\frac12\left(\nabla\phi_{\vec{x}}\right)^2+\frac12m^2\phi^2_{\vec{x}}+ig\phi^3_{\vec{x}}\right],
\end{equation}
where $\phi_{\vec{x}}\equiv\phi(\vec{x})$ and $\pi_{\vec{x}}\equiv\pi(\vec{x})$ are the Hermitian field variable and its canonical conjugate momentum satisfying
\begin{equation}
\mathcal{P}_I\phi_{\vec{x}}\mathcal{P}_I^{-1}=-\phi_{\vec{x}},\ \mathcal{P}_I\pi_{\vec{x}}\mathcal{P}_I^{-1}=-\pi_{\vec{x}},
\end{equation}
where $\mathcal{P}_I$ is the intrinsic parity operator. $Q$ in first order of $g$ is\cite{bender2005}
\begin{equation}\begin{aligned}
Q_1=\frac12g\int d^3xd^3yd^3z&\left[M(\vec{x},\vec{y},\vec{z})\pi_{\vec{x}}\pi_{\vec{y}}\pi_{\vec{z}}+N(\vec{x},\vec{y},\vec{z})\phi_{\vec{y}}\pi_{\vec{x}}\phi_{\vec{z}}\right],
\end{aligned}\end{equation}
where $M(\vec{x},\vec{y},\vec{z})$ and $N(\vec{x},\vec{y},\vec{z})$ are nonlocal functions. $\mathcal{C}$ is a lorentz scalar\cite{bender2005} but the above form of $\mathcal{C}=\mathcal{P}_Ie^{-2Q}$ is not manifestly lorentz invariant. In the next section we will develop the representation of $\mathcal{C}$ as well as $\eta$ in path integral formalism.

\section{Transformations in path integral formalism\label{sec-path}}

The Euclidean partition function of the form $\int D\phi e^{-S_E[\phi]}$ for $i\phi^3$ model is
\begin{equation}\label{eq-path}\begin{aligned}
&\int D\phi\exp\left\{-\int d^4x\left[\frac12\phi_x\left(-\partial^2_x+m^2\right)\phi_x+ig\phi^3_x\right]\right\},
\end{aligned}\end{equation}
where $\phi_x\equiv\phi(x)$ is a real pseudoscalar field.

We observe that under the transformation
\begin{equation}\label{eq-trans1}\begin{aligned}
\phi_x\rightarrow\rho\left(\phi_x\right)=&\phi_x-2ig\int d^4yD_{x-y}\phi^2_y\\
&-4g^2\int d^4yd^4zD_{x-y}\phi_yD_{y-z}\phi^2_z+\mathcal{O}(g^3),
\end{aligned}\end{equation}
where the Euclidean propagator $D_{x-y}\equiv D(x-y)=\int\frac{d^4p}{(2\pi)^4}\frac{e^{ip\cdot(x-y)}}{p^2+m^2}$ satisfies $\left(-\partial^2_x+m^2\right)D_{x-y}=\delta^4(x-y)$, the action transforms as
\begin{equation}
S_E[\phi]\rightarrow\int d^4x\left[\frac12\phi_x\left(-\partial^2+m^2\right)\phi_x-ig\phi^3_x\right]+\mathcal{O}(g^3).
\end{equation}
Assuming the contour of $\phi_x$ after transformation can be deformed to real axis without punishment, which is a nontrivial assumption but acceptable as long as we remain in perturbation theory, the action transforms into its complex conjugation up to order $g^3$.

The transformation (\ref{eq-trans1}) is simply a result from the completing-the-squrare procedure where $2ig\phi^3$ is absorbed into the quadratic term $\frac12\phi\left(-\partial^2+m^2\right)\phi$ order by order, and calculation of higher-order terms is systematic.

\begin{table}
\centering
\caption{\label{table-1}Relation of transformations of field between canonical and path integral formalism.}
\begin{tabular}{|c|c|c|}
\hline
canonical&path integral&relation to $\tau(\phi)$\\
\hline
$\tau\phi\tau^{-1}$&$\tau(\phi)$&$=\tau\left(\phi\right)$\\
\hline
$\tau^{-1}\phi\tau$&$\tau^{-1}(\phi)$&$=\tau^{-1}\left(\phi\right)$\\
\hline
$\tau^{-1\dagger}\phi\tau^\dagger$&$\tau^{-1\dagger}(\phi)$&$=\left[\tau\left(\phi\right)\right]^*$\\
\hline
$\tau^{\dagger}\phi\tau^{-1\dagger}$&$\tau^\dagger(\phi)$&$=\left[\tau^{-1}\left(\phi\right)\right]^*$\\
\hline
\end{tabular}
\end{table}

However, the measure in path integral also transforms, which serves as a source of quantum anomaly\cite{fujikawa}. The transformation (\ref{eq-trans1}) has a nontrivial Jacobian $\det\left[\delta\phi'_x/\delta\phi_y\right]$ calculated as follows
\begin{equation}\label{eq-anomaly}\begin{aligned}
\det&\left[\delta^4(x-y)-4igD_{x-y}\phi_y-4g^2\int d^4z D_{x-y}D_{y-z}\phi^2_z\right.\\
&\left.-8g^2\int d^4zD_{x-z}\phi_zD_{z-y}\phi_y+\mathcal{O}(g^3)\right]\\
=\exp&\left\{\mathrm{tr}\ \ln\left[\delta^4(x-y)-4igD_{x-y}\phi_y-4g^2\int d^4z D_{x-y}D_{y-z}\phi^2_z\right.\right.\\
&\left.\left.-8g^2\int d^4zD_{x-z}\phi_zD_{z-y}\phi_y+\mathcal{O}(g^3)\right]\right\}\\
=\exp&\left\{-\left[4ig\int d^4xD_0\phi_x+4g^2D_0\int d^4xd^4yD_{x-y}\phi^2_y+\mathcal{O}(g^3)\right]\right\}.
\end{aligned}\end{equation}
The measure thus contributes two anomalous terms to the action although appearing with obvious ultraviolet divergences which is predictable in 4-dimensional quantum field theory and can be dealt with using regularization methods such as dimensional regularization. To cancel these two terms, we add new terms in the transformation (\ref{eq-trans1})
\begin{equation}\label{eq-trans2}\begin{aligned}
\phi_x\rightarrow&\tilde{\rho}\left(\phi_x\right)=\phi_x-2ig\int d^4yD_{x-y}\phi^2_y-4g^2\int d^4yd^4zD_{x-y}\phi_yD_{y-z}\phi^2_z\\
&-4igD_0\int d^4yD_{x-y}-8g^2D_0\int d^4yd^4zD_{x-y}\phi_yD_{y-z}+\mathcal{O}(g^3),
\end{aligned}\end{equation}
such that anonamous terms from (\ref{eq-anomaly}) are cancelled out. There is a new anomalous term form the newly added order $g^2$ term, but it is cancelled out by the square of the newly added order $g$ term in the action because their effective contributions to the action are $\pm 8g^2D_0^2\int d^4xd^4yD_{x-y}$ respectively.

Clearly, the transformation (\ref{eq-trans2}) play the role of $\rho$ in canonical formalism and the difference is that the canonical $\rho$ transforms the Hamiltonian rather than the action to its Hermitian conjugation. If complemented by an intrinsic parity transformation $\phi_x\rightarrow-\phi_x$, the combined transformation
\begin{equation}\label{eq-c}\begin{aligned}
\phi_x\rightarrow&\mathcal{C}\left(\phi_x\right)=-\phi_x-2ig\int d^4yD_{x-y}\phi^2_y+4g^2\int d^4yd^4zD_{x-y}\phi_yD_{y-z}\phi^2_z\\
&-4igD_0\int d^4yD_{x-y}+8g^2D_0\int d^4yd^4zD_{x-y}\phi_yD_{y-z}+\mathcal{O}(g^3),
\end{aligned}\end{equation}
leaves the form of partition function invariant up to order $g^3$, which mimics the behavior of $\mathcal{C}$ operator in canonical formalism. It can be shown that $\mathcal{C}(\phi)$ commutes with $\mathcal{PT}(\alpha\phi)=-\alpha^*\phi$ for $\alpha\in\mathbbm{C}$ by observing the coefficients of separate terms.

We next solve for the transformation corresponding to $\eta$ using $\rho=\eta^\dagger\eta$ and the analogue between a transformation of field $\tau\phi\tau^{-1}$ in canonical formalism and a transformation $\phi\rightarrow\tau\left(\rho\right)$ in path integral formalism as shown in Table.\ref{table-1}.

Assume $\eta\left(\phi\right)$ is\footnote{The new term with coefficient $c_5$ is added because the anomalies of the $c_1$ and $c_2$ terms have contribution from $\int d^4d^4y \phi_xD_{x-y}^2\phi_y$ for generic values of $c_1$ and $c_2$ and it is necessary to add a term that generates the same term in the transformation of the action. Note that for $c_2=-c_1^2$, the anomalous terms proportional to $\int d^4xd^4y \phi_xD_{x-y}^2\phi_y$ produced by the $c_1$ and $c_2$ terms cancel with each other, which is exactly the case of (\ref{eq-trans2}).}
\begin{equation}\label{eq-eta}\begin{aligned}
\eta\left(\phi_x\right)=&\phi_x+c_1ig\int d^4yD_{x-y}\phi^2_y+c_2g^2\int d^4yd^4zD_{x-y}\phi_yD_{y-z}\phi^2_z\\
&+c_3ig D_0\int d^4yD_{x-y}+c_4g^2D_0\int d^4yd^4zD_{x-y}\phi_yD_{y-z}\\
&+c_5g^2\int d^4yd^4zD_{x-y}D_{y-z}^2\phi_z+\mathcal{O}(g^3).
\end{aligned}\end{equation}
Using Table.\ref{table-1}, $\eta^\dagger\left(\phi\right)$ is simply
\begin{equation}\begin{aligned}
\eta^\dagger\left(\phi_x\right)=&\phi_x+c_1ig\int d^4yD_{x-y}\phi^2_y-(c_2+2c_1^2)g^2\int d^4yd^4zD_{x-y}\phi_yD_{y-z}\phi^2_z\\
&+c_3ig D_0\int d^4yD_{x-y}-(2c_1c_3+c_4)g^2D_0\int d^4yd^4zD_{x-y}\phi_yD_{y-z}\\
&-c_5g^2\int d^4yd^4zD_{x-y}D_{y-z}^2\phi_z+\mathcal{O}(g^3),
\end{aligned}\end{equation}
which fulfills the relation $\eta^\dagger(\phi)=\left[\eta^{-1}(\phi)\right]^*$. The relation $\tilde{\rho}\left(\phi_x\right)=\left(\eta^\dagger\eta\right)\left(\phi_x\right)=\eta^\dagger\left(\eta\left(\phi_x\right)\right)$ gives constraint equation as follows
\begin{equation}
2c_1=-2,\ -4c_1^2=-4,\ 2c_3=-4,\ -4c_1c_3=-8,
\end{equation}
which requires $c_1=-1$ and $c_3=-2$, but does not constrain $c_2$, $c_4$ and $c_5$. 

Before going to the next section, we want to show that the method presented in this section to derive $\mathcal{C}$ and $\eta$ is a natural generalization of the completing-the-square method in the free $\mathcal{PT}$-symmetric $i\phi$ model which is almost the simplest $\mathcal{PT}$-symmetric model and used as a toy model\cite{bender2005}. For the partition function $\int D\phi\exp\left\{-\int d^4x\left[\frac12\phi_x\left(-\partial^2_x+m^2\right)\phi_x+ig\phi_x\right]\right\}$, the corresponding Hermitian theory can be derived by completing the square in the action as follows
\begin{equation}\begin{aligned}
&\int d^4x\left[\frac12\phi_x\left(-\partial^2_x+m^2\right)\phi_x+ig\phi_x\right]\\
=&\int d^4x\left[\frac12\left(\phi_x+ig\int d^4yD_{x-y}\right)\left(-\partial^2_x+m^2\right)\left(\phi_x+ig\int d^4zD_{x-z}\right)\right.\\
&\left.+\frac12g^2\int d^4yd^4zD_{y-z}\right],
\end{aligned}\end{equation}
such that $\eta(\phi_x)=\phi_x-ig\int d^4yD_{x-y}=\phi_x-\frac{ig}{m^2}$ in the $i\phi$ model. It is apparent that the method in this section is a natural continuation of the completing-the-square method from the free model to interacting models while the series are not truncated in finite orders any more and obtain corrections from anomalies.

\section{Infinite Physically Inequivalent Hermitian Theories\label{sec-infinite}}

The transformed Euclidean action under (\ref{eq-eta}) is
\begin{equation}\begin{aligned}
&\int d^4x\frac12\phi_x\left(-\partial^2_x+m^2\right)\phi_x+\left(c_2+\frac52\right)g^2\int d^4xd^4y\phi^2_xD_{x-y}\phi^2_y\\
&-2igD(0)\int d^4x\phi_x-2g^2D_0^2\int d^4xd^4yD_{x-y}\\
&+(4+c_4)g^2D_0\int d^4xd^4yD_{x-y}\phi^2_y+c_5g^2\int d^4xd^4y\phi_xD_{x-y}^2\phi_y+\mathcal{O}(g^3).
\end{aligned}\end{equation}
The effective contribution of anomaly from (\ref{eq-eta}) to the action is 
\begin{equation}\begin{aligned}
-&\left[-2igD_0\int d^4x\phi_x+c_2g^2D_0\int d^4xd^4yD_{x-y}\phi^2_y\right.\\
&+(2c_2+2)g^2\int d^4xd^4y\phi_xD_{x-y}^2\phi_y+c_4g^2D_0^2\int d^4xd^4yD_{x-y}\\
&\left.+c_5g^2\int d^4xd^4yD_{x-y}^3\right]+\mathcal{O}(g^3).
\end{aligned}\end{equation}
Therefore, the full result for the transformed partition function is
\begin{equation}\label{eq-equiv}\begin{aligned}
&\int D\phi\exp\left\{-\left[\int d^4x\frac12\phi_x\left(-\partial^2_x+m^2\right)\phi_x+\left(c_2+\frac52\right)g^2\int d^4xd^4y\phi^2_xD_{x-y}\phi^2_y\right.\right.\\
&-(2+c_4)g^2D_0^2\int d^4xd^4yD_{x-y}+(4+c_4-c_2)g^2D_0\int d^4xd^4yD_{x-y}\phi^2_y\\
&\left.\left.+(c_5-2c_2-2)g^2\int d^4xd^4y\phi_xD_{x-y}^2\phi_y-c_5g^2\int d^4xd^4yD_{x-y}^3+\mathcal{O}(g^3)\right]\right\}.
\end{aligned}\end{equation}
Because $c_2$, $c_4$ and $c_5$ are not constrained, the corresponding Hermitian theory for $\mathcal{PT}$-symmetric $i\phi^3$ is arbitrary. In particular, if $c_2=-\frac52$, the functional integral (\ref{eq-equiv}) is quadratic up to order $g^3$ such that the $S$ matrix obtained by LSZ formula after Wick rotation is trivial.

To conclude the arbitrariness of the $S$ matrix, there need to be some explanations. The functional integral (\ref{eq-path}), or (\ref{eq-equiv}), is actually independent of field redefinitions because it is an integral without any external variables. To calculate the $S$ matrix from Green's functions, external sources must be added into the functional integrals. However, it is the Green's functions with physical fields that lead to unitary $S$ matrices. The field variable $\phi_x$ in the original $i\phi^3$ path integral (\ref{eq-path}) is not a physical field and corresponding $S$ matrix by the LSZ formula is not unitary because the $i\phi^3$ action is not Hermitian. Therefore, $-\int d^4x \phi^{\mathrm{phys}}_x j_x$ rather than $-\int d^4x \phi_x j_x$  should be added to the exponent in (\ref{eq-path}), where $j_x$ is the external source and the $i\phi^3$ action should be transformed to a Hermitian one under the transformation $\phi^{\mathrm{phys}}_x\rightarrow \phi_x$ as explained in \cite{bender-book,jones2007}. Under the $\eta$ transformation (\ref{eq-eta}), the $i\phi^3$ action becomes Hermitian as shown by (\ref{eq-equiv}), and thus $\phi^{\mathrm{phys}}_x$ can be chosen to be $\eta^{-1}(\phi_x)$ resulting a $-\int d^4x \phi_x j_x$ term in the exponent of (\ref{eq-equiv}). Different $\eta$ transformations lead to different choices of the physical field $\phi^{\mathrm{phys}}_x$. Supplemented with a source term, the arbitrariness of the coefficients of terms in (\ref{eq-equiv}) leads to different theories with different Green's functions resulting different $S$ matrices by the LSZ formula.

The above shows that nonlocal field redefinitions lead to physically unacceptable results for $\mathcal{PT}$-symmetric $i\phi^3$ model. And we will show this is also the case for Hermitian theories such as $\lambda\phi^4$ model with $\lambda>0$.

\section{Nonlocal Transformations in Hermitian Theories\label{sec-herm}}

Consider the change of the Euclidean partition function for $\lambda\phi^4$
\begin{equation}\begin{aligned}
\int D\phi\exp\left\{-\int d^4x\left[\frac12\phi_x\left(-\partial^2_x+m^2\right)\phi_x+\lambda\phi^4_x\right]\right\},
\end{aligned}\end{equation}
under the transformation as follows
\begin{equation}\begin{aligned}
\phi_x\rightarrow F\left(\phi_x\right)=&\phi_x+f_1\lambda\int d^4yD_{x-y}\phi^3_y+f_2\lambda D_0\int d^4yD_{x-y}\phi_y+\mathcal{O}(\lambda^2).
\end{aligned}\end{equation}
The transformed partition function is
\begin{equation}\begin{aligned}
\int D\phi\exp&\left\{-\int d^4x\left[\frac12\phi_x\left(-\partial^2_x+m^2\right)\phi_x+\left(1+f_1\right)\lambda\phi^4_x\right.\right.\\
&\left.\left.+\left(f_2-3f_1\right)\lambda D_0\phi^2_x-f_2\lambda D_0^2\right]+\mathcal{O}(\lambda^2)\right\}.
\end{aligned}\end{equation}
Clearly there are vast degrees of freedom to adjust the transformed theory.

However, the $\lambda\phi^4$ action is already Hermitian without making use of any transformation. Hermitian local actions are sufficient to generate perturbative unitarity and causality\cite{weinberg}, so there is no need to consider the above nonlocal transformations. It is not the case in $\mathcal{PT}$-symmetric theories, because a non-Hermitian local action cannot lead to a unitary $S$ matrix without transforming to a Hermitian one, and these transformations are shown to be nonlocal in both canonical formalism such as \cite{bender2004a} and path integral formalism as in this paper. So it is an inevitable step to deal with the arbitrariness problem before $\mathcal{PT}$-symmetric theories can be put into use.

\section{$0$-dimensional Models\label{sec-0}}

We have discussed perturbation theories only, and in this section we give some nonperturbative arguments about the $\mathcal{C}=\mathcal{P}\tilde{\rho}$ and $\eta$ transformations. The $\tilde{\rho}$ transformation (\ref{eq-trans2}) is actually solved from the equation
\begin{equation}\label{eq-nonp}\begin{aligned}
&\int D\phi\det\left[\frac{\delta\rho(\phi_x)}{\delta\phi_y}\right]\exp\left\{\int d^4x\left[\frac12\tilde{\rho}(\phi_x)\left(-\partial^2+m^2\right)\tilde{\rho}(\phi_x)+ig\tilde{\rho}(\phi_x)^3\right]\right\}\\
=&\int D\phi\exp\left\{\int d^4x\left[\frac12\phi_x\left(-\partial^2+m^2\right)\phi_x-ig\phi^3_x\right]\right\},
\end{aligned}\end{equation}
where $\tilde{\rho}$ transforms the action to its Hermitian conjugation. And (\ref{eq-trans2}) actually gives a perturbative expansion by the completing-the-square procedure. (\ref{eq-nonp}) is difficult to solve analytically, and we turn to $0$-dimensional $i\phi^3$ to obtain some nonperturbative properties of the $\tilde{\rho}$ transformation numerically.

\begin{figure}[ht]
\centering
\begin{minipage}{0.49\linewidth}
\includegraphics[width=0.9\linewidth]{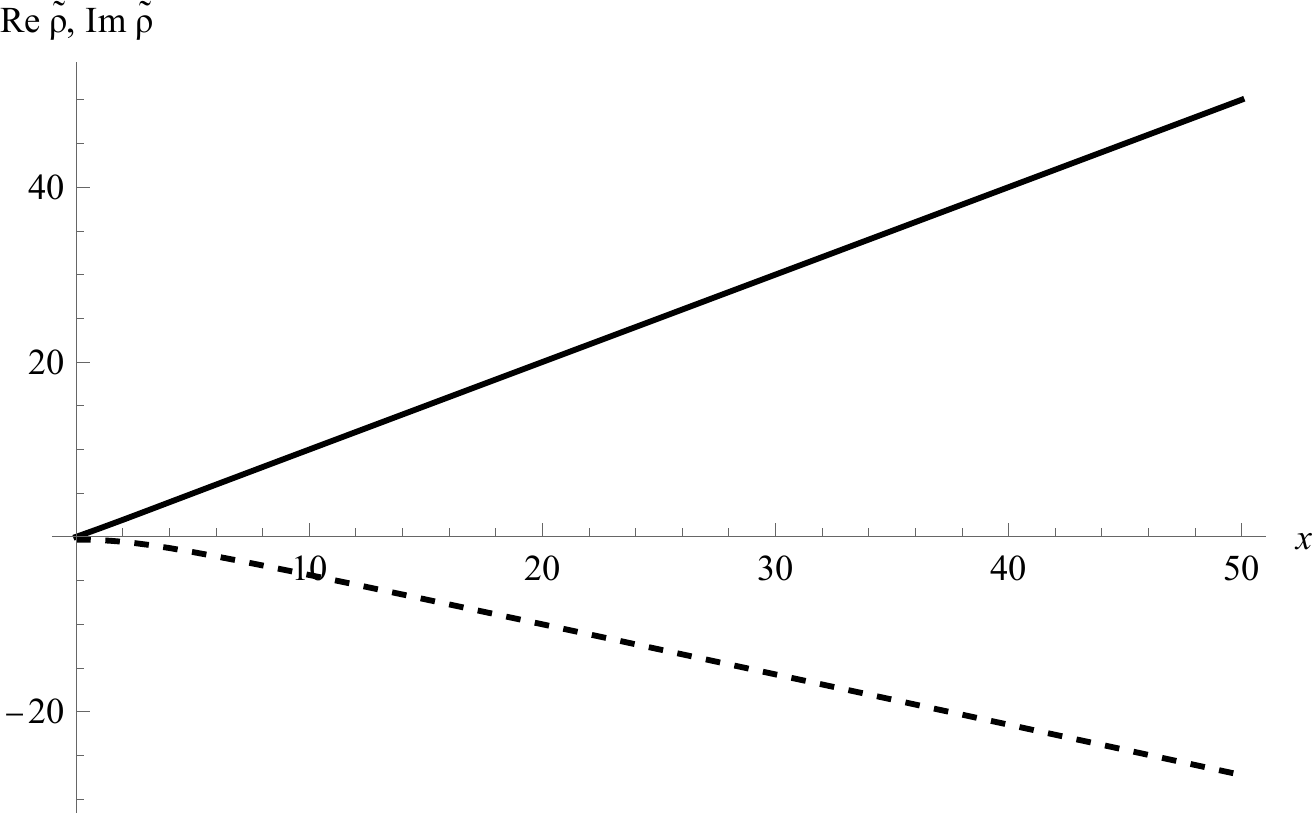}
\caption{Numerical results of $\tilde{\rho}(x)$ with $m=1,g=0.1,x\in\left[0,50\right]$. $\mathrm{Re}\tilde{\rho}(x)$ and $\mathrm{Im}\tilde{\rho}(x)$ are represented by solid and dashed lines, respectively.}
\label{fig-1}
\end{minipage}
\begin{minipage}{0.49\linewidth}
\includegraphics[width=0.9\linewidth]{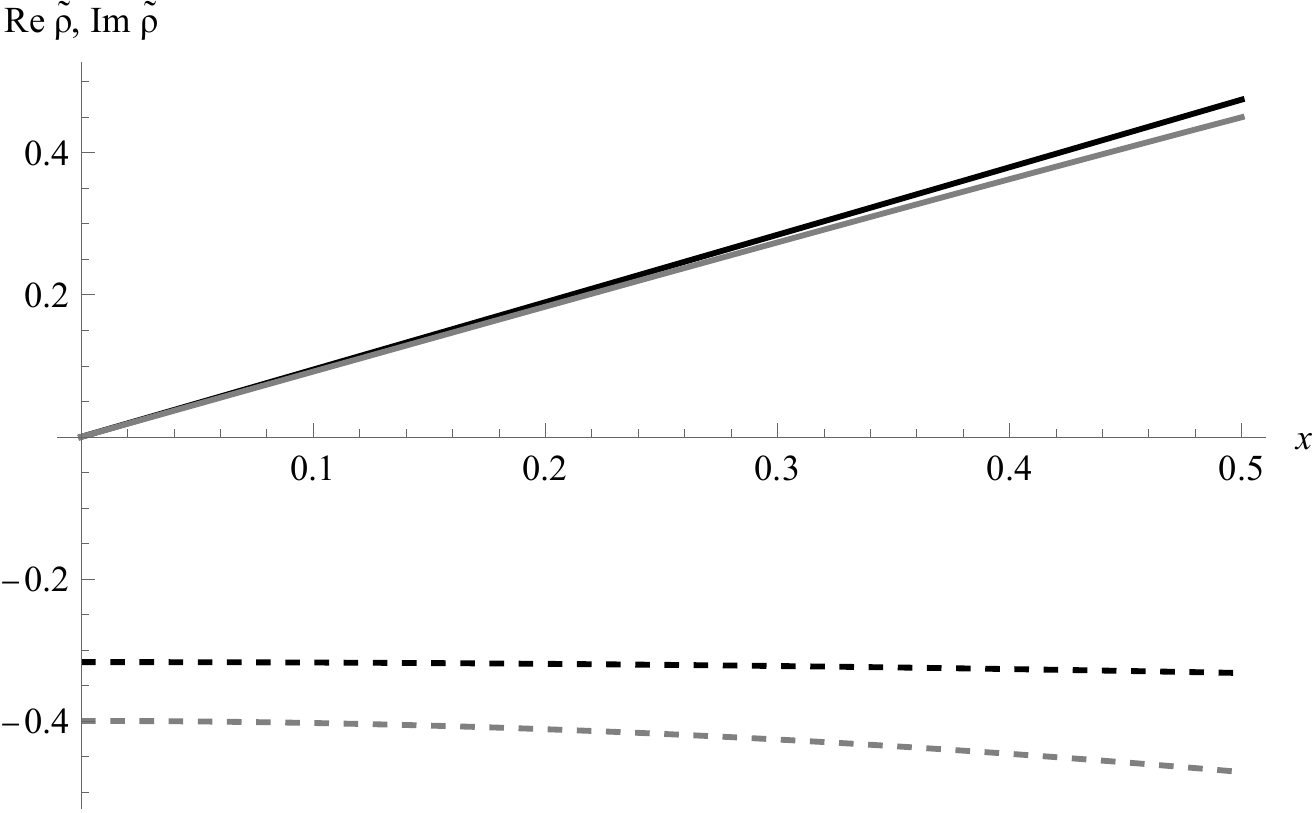}
\caption{$x\in\left[0,0.5\right]$ region of Fig. \ref{fig-1}, with perturbative results of $\mathrm{Re}\tilde{\rho}(x)$ and $\mathrm{Im}\tilde{\rho}(x)$ in gray solid and dashed lines.}
\label{fig-2}
\end{minipage}
\end{figure}

The path integral representation of $0$-dimensional $i\phi^3$ is
\begin{equation}\label{eq-0z}
Z_g=\int_C d\phi \exp\left(-\frac12m^2\phi^2-ig\phi^3\right),
\end{equation}
where $C$ can be any contour in the Stokes sectors where real axis lives\cite{bender-book}. The $0$-dimensional $\tilde{\rho}$ transformation is therefore the solution of
\begin{equation}\label{eq-01}\begin{aligned}
\int_C d\phi\tilde{\rho}'(\phi)\exp\left(-\frac12m^2\tilde{\rho}(\phi)^2-ig\tilde{\rho}(\phi)^3\right)=\int_{\bar{C}}d\phi\exp\left(-\frac12m^2\phi^2+ig\phi^3\right),
\end{aligned}\end{equation}
where $\bar{C}$ is the complex conjugate contour of $C$. Without the integration, (\ref{eq-01}) is a first order ordinary differential equation, and one boundary condition should be specified to obtian a unique solution. We expect $\tilde{\rho}(\phi)$ lies in a contour in the same Stokes sector as that before transformation, and thus the behavior of $\tilde{\rho}(\phi)$ when $|\phi|\rightarrow\infty$ can serve as the boundary condition. To improve the convergence of numerical algorithm, we take $C$ as the constant phase contour\cite{bender-book} and parametrize the field variable as
\begin{equation}\label{eq-02}\begin{aligned}
\phi_C(x)=x+i\frac{m^2-\sqrt{m^4+12g^2x^2}}{6g},
\end{aligned}\end{equation}
where $x\in\mathbbm{R}$. Along the contour (\ref{eq-02}), the argument of the exponent in(\ref{eq-0z}) is always real. The contour $\bar{C}$ is thus parametrized as
\begin{equation}\label{eq-03}\begin{aligned}
\phi_{\bar{C}}(x)=x-i\frac{m^2-\sqrt{m^4+12g^2x^2}}{6g},
\end{aligned}\end{equation}
which is the complex conjugation of (\ref{eq-02}). Using the contour (\ref{eq-02}) and (\ref{eq-03}), the equation and boundary condition for $\tilde{\rho}(x)\equiv\tilde{\rho}(\phi_C(x))$ can be written down immediately as follows,
\begin{equation}\label{eq-04}\begin{aligned}
&\tilde{\rho}'(x)\exp\left(-\frac12m^2\tilde{\rho}(x)^2-ig\tilde{\rho}(x)^3\right)=\phi'_{\bar{C}}(x)\left.\exp\left(-\frac12m^2\phi^2+ig\phi^3\right)\right|_{\phi=\phi_{\bar{C}}(x)}, \\
&\tilde{\rho}(x)\rightarrow\phi_C(x)\mbox{ when }|x|\rightarrow\infty.
\end{aligned}\end{equation}
This equation exhibits $\mathcal{PT}$ symmetry resulting $\tilde{\rho}(x)^*=-\tilde{\rho}(-x)$, and we only have to numerically integrate $\tilde{\rho}(x)$ on half of real axis. An elementary numerical algorithm gives the solution of $\tilde{\rho}(x)$ as shown in Fig. \ref{fig-1} and Fig. \ref{fig-2}. The vanishment of $\mathrm{Re}\tilde{\rho}(0)$ indicates $\mathcal{PT}$ symmetry is preserved by the numerical algorithm. We can also compare the numerical result with the perturbative result of $\tilde{\rho}(x)$ which is
\begin{equation}\label{eq-05}\begin{aligned}
\tilde{\rho}(x)=\left.\phi-2i\frac{g}{m^2}\phi^2-4\frac{g^2}{m^4}\phi^3-4i\frac{g}{m^2}-8\frac{g^2}{m^6}\phi\right|_{\phi=\phi_C(x)},
\end{aligned}\end{equation}
making use of (\ref{eq-trans2}) because coefficients in (\ref{eq-trans2}) is independent of dimension. This is shown in Fig. \ref{fig-2}, and the difference between numerical and perturbative results is of higher order\footnote{Although perturbative $\mathrm{Im}\tilde{\rho}(x)$ seems to differ a lot from numerical results, it has the same quality as Feynman diagramatic perturbation theories. For example, when $m=1,g=0.1$, the two point function is
\[\begin{aligned}
G_2=&\left[\int d\phi e^{-m^2\phi^2/2-ig\phi^3}\right]^{-1}\int d\phi e^{-m^2\phi^2/2-ig\phi^3}\phi^2\\
=&0.755810,
\end{aligned}\]
and 1-loop Feynman diagrams give perturbative result
\[\begin{aligned}
G_2^{(1)}=&\left[\int d\phi e^{-m^2\phi^2/2}+\int d\phi e^{-m^2\phi^2/2}(-ig\phi^3)^2/2\right]^{-1}\\
&\times\left[\int d\phi e^{-m^2\phi^2}\phi^2+\int d\phi e^{-m^2\phi^2/2}\phi^2(-ig\phi^3)^2/2\right]\\
=&0.513514.
\end{aligned}\]
Obviously, our perturbative result of $\mathrm{Im}\tilde{\rho}(x)$ is not bad comparing to Feynman diagramatic perturbation theories.}. Of course, the perturbative result breaks down in large $x$, but the validity of perturbative expansion is guaranteed by the existence of the numerical result.

\begin{figure}[ht]
\centering
\begin{minipage}{0.49\linewidth}
\includegraphics[width=0.9\linewidth]{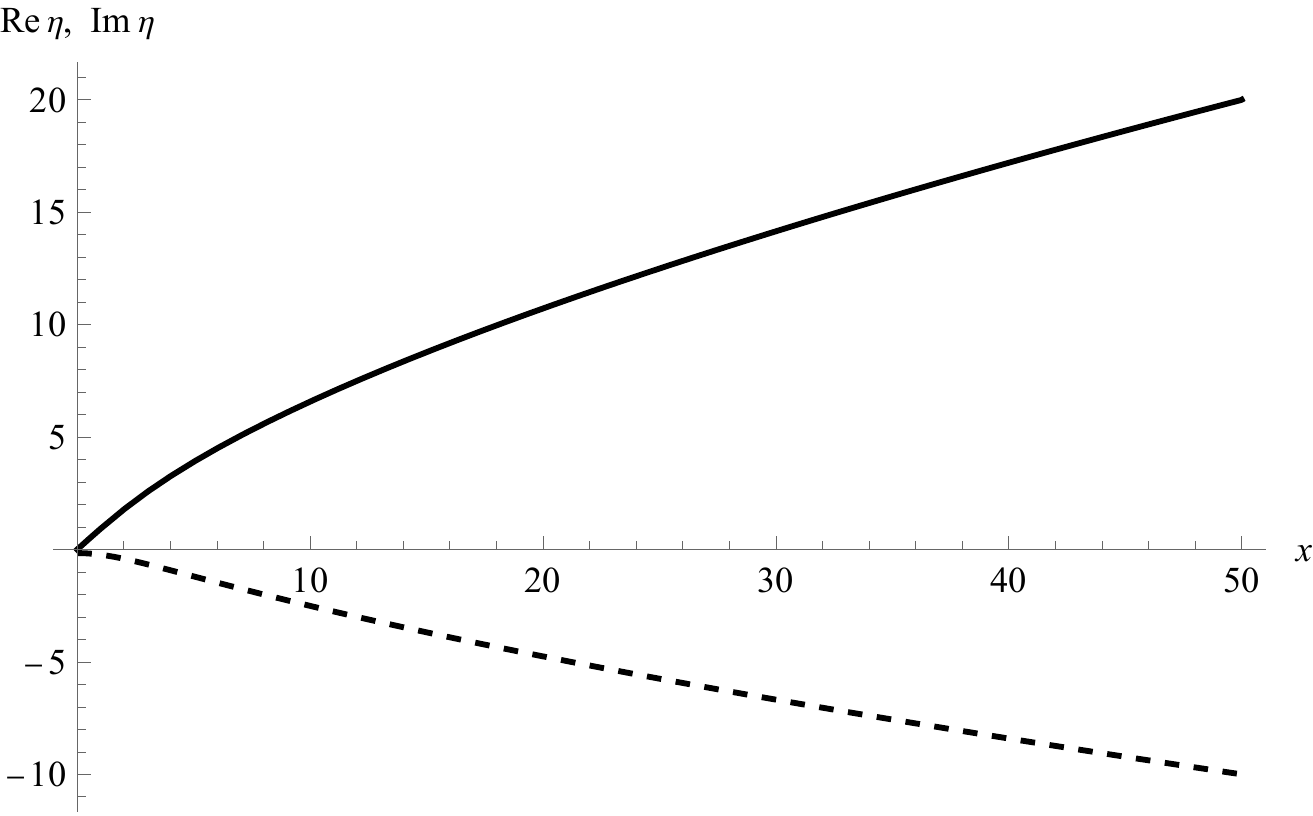}
\caption{Numerical results of $\eta(x)$ with $m=1,g=0.1,x\in\left[0,50\right]$. $\Delta E_0$ is found to be $0.053405$. $\mathrm{Re}\eta(x)$ and $\mathrm{Im}\eta(x)$ are represented by solid and dashed lines, respectively.}
\label{fig-3}
\end{minipage}
\begin{minipage}{0.49\linewidth}
\includegraphics[width=0.9\linewidth]{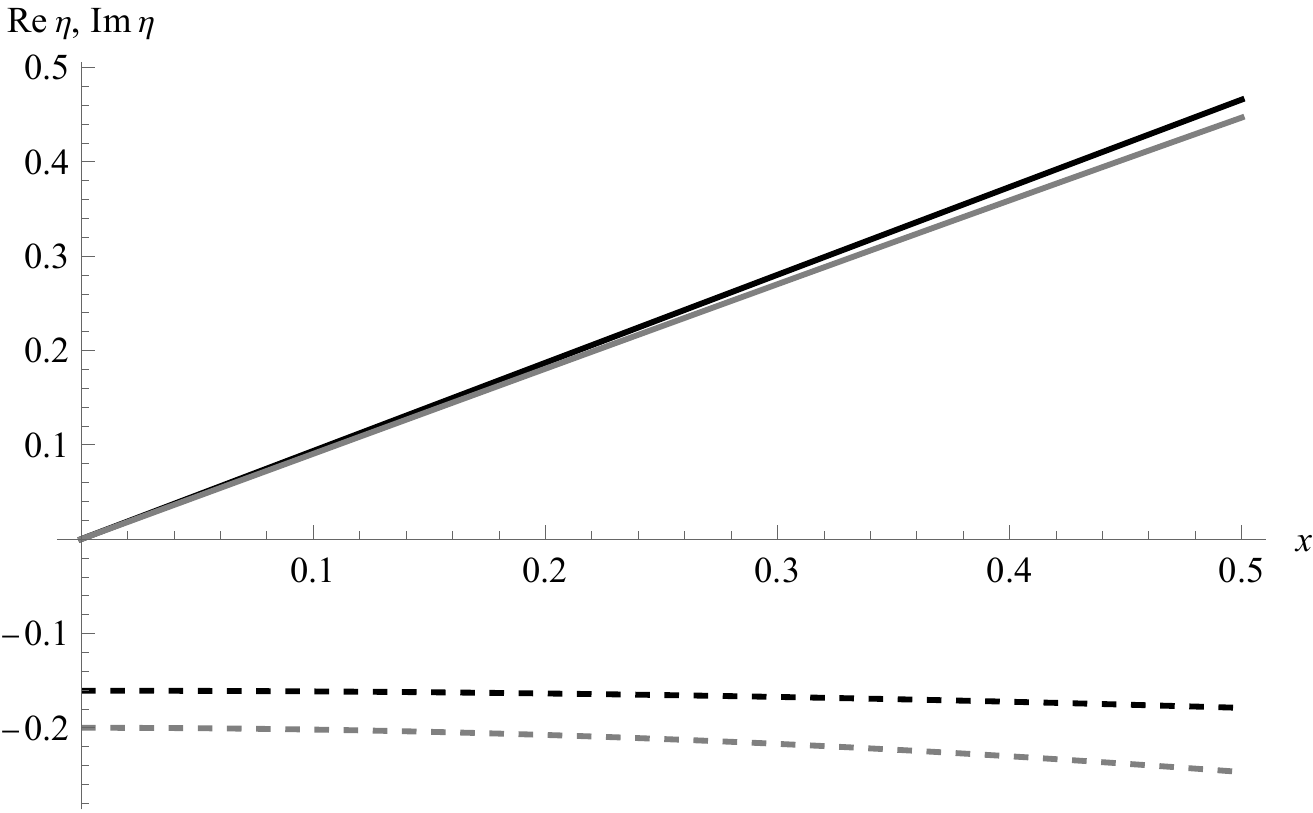}
\caption{$x\in\left[0,0.5\right]$ region of Fig. \ref{fig-3}, with perturbative results of $\mathrm{Re}\eta(x)$ and $\mathrm{Im}\eta(x)$ in gray solid and dashed lines.}
\label{fig-4}
\end{minipage}
\end{figure}

Having established the existence of the $\tilde{\rho}$ tranformation, we turn to the $\eta$ transformation. A remarkable feature of the $\eta$ transformation in (\ref{eq-eta}) is that the transformed action can be quadratic and thus has no interaction, and here we examine this special case. The $\eta$ transformation is solved from the following equation
\begin{equation}\label{eq-06}\begin{aligned}
\int_C d\phi\eta'(\phi)\exp\left(-\frac12m^2\eta(\phi)^2-ig\eta(\phi)^3\right)=\int_{\mathbbm{R}}d\phi\exp\left(-\frac12m^2\phi^2-\Delta E_0\right).
\end{aligned}\end{equation}
Taking $C$ to be the constant phase contour (\ref{eq-02}) again, we can immediately write down the equation satisfied by $\eta(x)\equiv\eta(\phi_C(x))$ as follows,
\begin{equation}\label{eq-07}\begin{aligned}
&\eta'(x)\exp\left(-\frac12m^2\eta(x)^2-ig\eta(x)^3\right)=\exp\left(-\frac12m^2x^2-\Delta E_0\right),\\
&\frac12m^2\eta(x)^2+ig\eta(x)^3\rightarrow\frac12m^2x^2\mbox{ when }|x|\rightarrow\infty,
\end{aligned}\end{equation}
where the boundary condition simply states $\eta'(x)$ is subleading which is reasonable beacuse we expect the transformed contour $\eta(x)$ varies relatively slowly in large $x$, and $\Delta E_0$ can be decided by the $\mathcal{PT}$ symmety of $\eta(x)$ which requires $\eta^*(x)=-\eta(-x)$. Numerical result of $\eta(x)$ is shown in Fig. \ref{fig-3} and Fig. \ref{fig-4}. Coefficients in (\ref{eq-eta}) that transforms $i\phi^3$ to a free theory is $c_1=-1,c_2=-5/2,c_3=-2,c_4=-13/2,c_5=-5$, and corresponding $0$-dimensional perturbative results of $\eta(x)$ is thus
\begin{equation}\label{eq-08}\begin{aligned}
\eta(x)=\left.\phi-i\frac{g}{m^2}\phi^2-\frac52\frac{g^2}{m^4}\phi^3-2i\frac{g}{m^4}-\frac{19}{2}\frac{g^2}{m^6}\phi\right|_{\phi=\phi_C(x)},
\end{aligned}\end{equation}
which is compared with numerical results in Fig. \ref{fig-4}, and the difference between numerical and perturbative results is of higher order. (\ref{eq-equiv}) also predicts $\Delta E_0=(15g^2)/(2m^6)=0.075$ in the numerical setting, which gives leading contribution comparing to the numerical result $0.053405$.

\begin{figure}[ht]
\centering
\begin{minipage}{0.49\linewidth}
\includegraphics[width=0.9\linewidth]{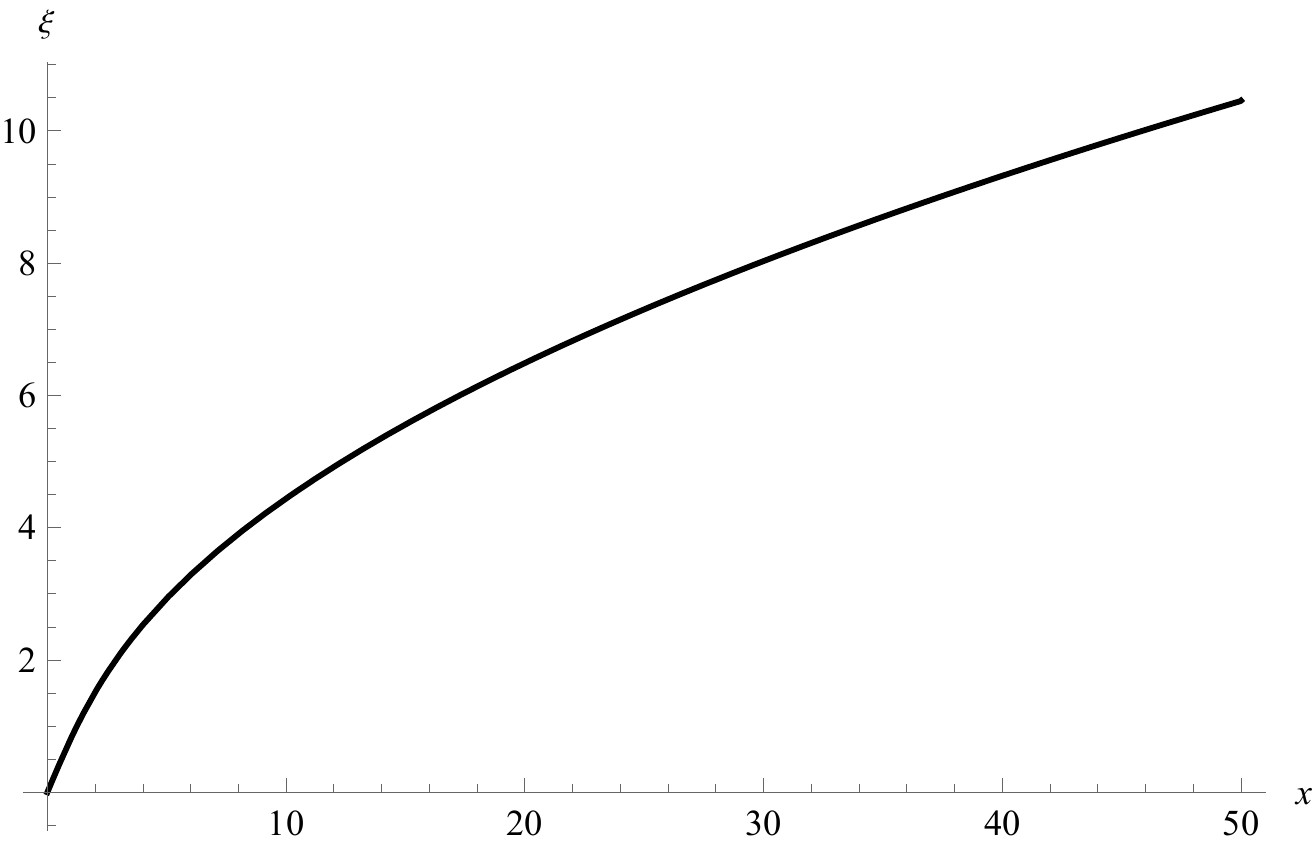}
\caption{Numerical results of $\xi(x)$ with $m=1,V(x)=0.1x^4,x\in\left[0,50\right]$. $\Delta E_0$ is found to be $0.153608$.}
\label{fig-5}
\end{minipage}
\begin{minipage}{0.49\linewidth}
\includegraphics[width=0.9\linewidth]{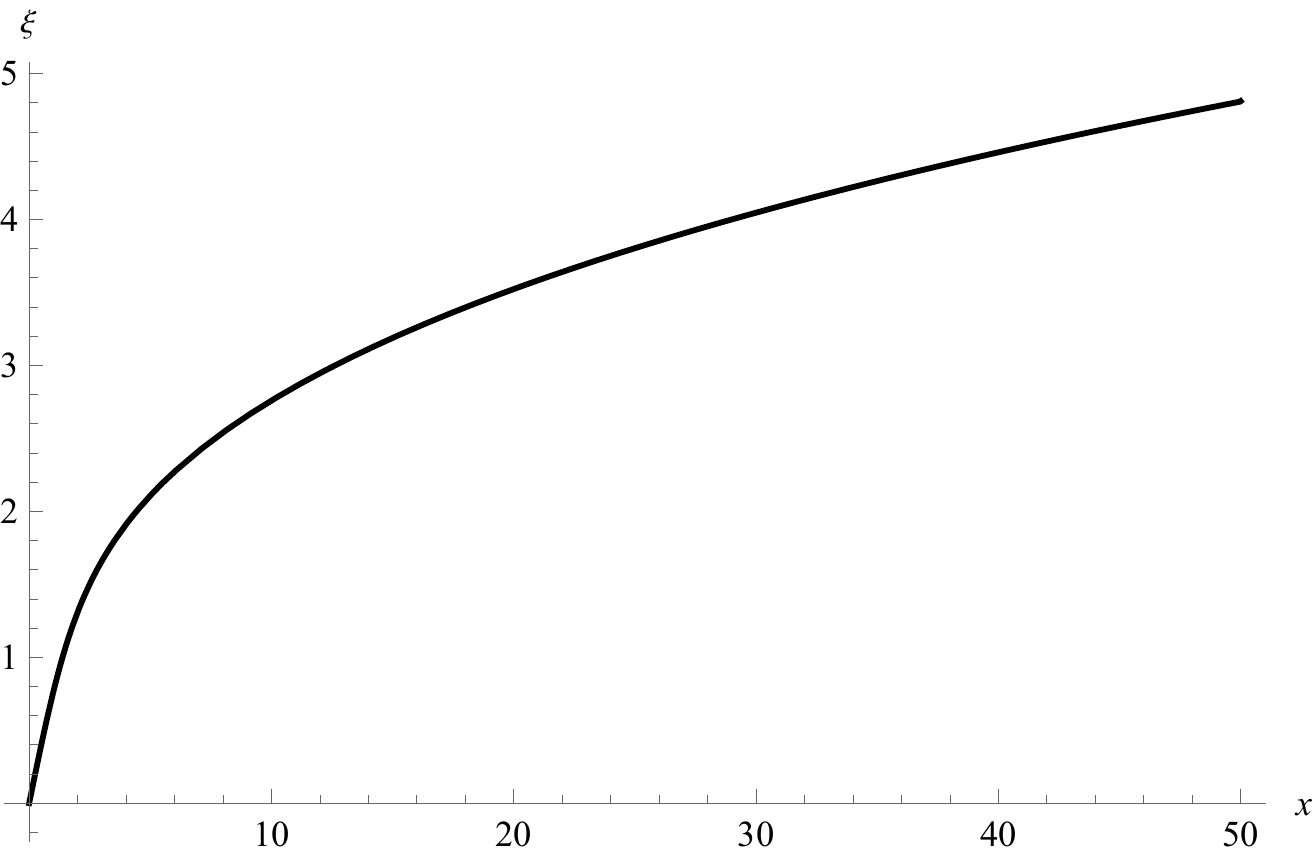}
\caption{Numerical results of $\xi(x)$ with $m=1,V(x)=0.1x^6,x\in\left[0,50\right]$. $\Delta E_0$ is found to be $0.208520$.}
\label{fig-6}
\end{minipage}
\end{figure}

Although weird, the transformation from $i\phi^3$ model to a quadratic model does exist, at least in $0$-dimension. And we futher show Hermitian non-quadratic actions can also be transformed to quadratic actions in $0$-dimension, by explicitly solving the following equation numerically for $V(x)=\lambda x^4,\lambda x^6$,
\begin{equation}\label{eq-09}\begin{aligned}
&\xi'(x)\exp\left(-\frac12m^2\xi(x)^2-V(\xi(x))\right)=\exp\left(-\frac12m^2x^2-\Delta E_0\right),\\
&\frac12m^2\xi(x)^2-V(\xi(x))\rightarrow\frac12m^2x^2\mbox{ when }|x|\rightarrow\infty,
\end{aligned}\end{equation}
where $\xi(x)$ is the corresponding tranformation. And $\Delta E_0$ can still be decided by the remaing $\mathcal{PT}$ symmety of $\xi(x)$ which requires $\xi(x)=-\xi(-x)$. Numerical results are shown in Fig. \ref{fig-5} and Fig. \ref{fig-6}.

\section{Conclusions\label{sec-final}}

We have reformulated the $\mathcal{C}$ and $\eta$ operators related by $\mathcal{C}=\mathcal{P}\eta^\dagger\eta$, which are keys to calculate physical quantities in the $\mathcal{PT}$-symmetric $i\phi^3$ model, as manifestly 4-dimensional invariant transformations of field variables in path integral formalism. It is found that the transformation $\eta(\phi)$ as in (\ref{eq-eta}) is not unique and transforms $i\phi^3$ to infinite Hermitian theories as shown by (\ref{eq-equiv}). No obvious arguments can be made to constrain parameters in the transformation. The existence of any physical meaning of perturbative $\mathcal{PT}$-symmetric models such as $i\phi^3$ model is thus a severe problem that prevents $\mathcal{PT}$-symmetric theories to be used in realistic model building.

To find the way out, we review possible deficiencies in this work. First, the convergence of perturbation series such as (\ref{eq-c}) and (\ref{eq-eta}) is not guaranteed in $4$-dimension, and non-perturbative corrections may be of great importance. Second, we assume the contour of transformed fields can be deformed back to real axis, which is to say transformations proposed in this paper do not alter the Stokes sectors where the fields live thus lead to the same functional integral\cite{bender-book}, and non-perturbative arguments in $4$-dimension are also needed to show this is indeed the case. Finally, we calculate only to the second order in the $i\phi^3$ model and higher-order results may be nontrivial and put constraints on $\eta(\phi)$. However, higher-order calculation is much more complicated and will be studied  in future works. But at least, we have shown $0$-dimensional $i\phi^3$ model behaves well in the sense that transformations representing $C$ and $\eta$ operators do exist numerically and match perturbative expansions in small-coupling and small-field regions, which should give a primary confirmation of our proposal, noting that coefficients of those transformations are independent of dimension.

Moreover, in \cite{david2023} the global $U(1)$ parameter from a gauge theory was promoted to a complex number thus resulting $\mathcal{PT}$-symmetric Hamiltonians, and the arbitrariness of the $\eta$ operator is compensated by the gauge invariance of the theory. This suggests finding hidden symmetries in $\mathcal{PT}$-symmetric theories to reduce the large degrees of freedom of the $\eta$ operator. Optimistically, if the choice can be made to select the unique transformation that carries $i\phi^3$ to its Hermitian counterpart, the representation of the transformation in path integral formalism has another advantage comparing to canonical formalism besides the manifest 4-dimensional invariance, which is the manifest compatability with causality. After Wick-rotating to Minkowski space, the Euclidean propagator $D(x-y)=\int\frac{d^4p}{(2\pi)^4}\frac{e^{ip\cdot(x-y)}}{p^2+m^2}$ is simply substituted by the Feynman propagator $D_F(x-y)=\int \frac{d^4p}{(2\pi^4)}\frac{ie^{-ip\cdot(x-y)}}{p^2-m^2+i0_+}$, and the causality violation in canonical formalism owing to the presence of non-Feynman propagator described in \cite{novikov2019} can be saved immediately. 

\section*{Acknowledgements}
We thank Qi Chen, Zi-Kan Geng, Wei-Jun Kong, Yu-Hang Li and Chen Yang for inspiring discussions.

\bibliographystyle{elsarticle-harv} 
\bibliography{ref.bib}

\begin{thebibliography}{29}
\expandafter\ifx\csname natexlab\endcsname\relax\def\natexlab#1{#1}\fi
\providecommand{\url}[1]{\texttt{#1}}
\providecommand{\href}[2]{#2}
\providecommand{\path}[1]{#1}
\providecommand{\DOIprefix}{doi:}
\providecommand{\ArXivprefix}{arXiv:}
\providecommand{\URLprefix}{URL: }
\providecommand{\Pubmedprefix}{pmid:}
\providecommand{\doi}[1]{\href{http://dx.doi.org/#1}{\path{#1}}}
\providecommand{\Pubmed}[1]{\href{pmid:#1}{\path{#1}}}
\providecommand{\bibinfo}[2]{#2}
\ifx\xfnm\relax \def\xfnm[#1]{\unskip,\space#1}\fi
\bibitem[{Bender and Boettcher(1998)}]{bender1998}
\bibinfo{author}{Bender, C.}, \bibinfo{author}{Boettcher, S.},
  \bibinfo{year}{1998}.
\newblock \bibinfo{title}{Real spectra in non-hermitian hamiltonians having pt
  symmetry}.
\newblock \bibinfo{journal}{Phys. Rev. Lett.} \bibinfo{volume}{80},
  \bibinfo{pages}{5243--5246}.
\newblock \DOIprefix\doi{10.1103/PhysRevLett.80.5243}.
\bibitem[{Bender(2007)}]{bender-review}
\bibinfo{author}{Bender, C.M.}, \bibinfo{year}{2007}.
\newblock \bibinfo{title}{Making sense of non-hermitian hamiltonians}.
\newblock \bibinfo{journal}{Rep. Prog. Phys.} \bibinfo{volume}{70},
  \bibinfo{pages}{947}.
\newblock \DOIprefix\doi{10.1088/0034-4885/70/6/R03}.
\bibitem[{Bender(2011)}]{bender2011}
\bibinfo{author}{Bender, C.M.}, \bibinfo{year}{2011}.
\newblock \bibinfo{title}{Pt‐symmetric quantum field theory}.
\newblock \bibinfo{journal}{AIP Conference Proceedings} \bibinfo{volume}{1389},
  \bibinfo{pages}{642--645}.
\newblock \DOIprefix\doi{10.1063/1.3636813}.
\bibitem[{Bender et~al.(2002a)Bender, Berry and Mandilara}]{bender2002-gen}
\bibinfo{author}{Bender, C.M.}, \bibinfo{author}{Berry, M.V.},
  \bibinfo{author}{Mandilara, A.}, \bibinfo{year}{2002}a.
\newblock \bibinfo{title}{Generalized pt symmetry and real spectra}.
\newblock \bibinfo{journal}{J. Phys. A: Math. Gen.} \bibinfo{volume}{35},
  \bibinfo{pages}{L467}.
\newblock \DOIprefix\doi{10.1088/0305-4470/35/31/101}.
\bibitem[{Bender et~al.(2012)Bender, Branchina and Messina}]{bender2012b}
\bibinfo{author}{Bender, C.M.}, \bibinfo{author}{Branchina, V.},
  \bibinfo{author}{Messina, E.}, \bibinfo{year}{2012}.
\newblock \bibinfo{title}{Ordinary versus $\mathcal{P}\mathcal{T}$-symmetric
  ${\phi}^{3}$ quantum field theory}.
\newblock \bibinfo{journal}{Phys. Rev. D} \bibinfo{volume}{85},
  \bibinfo{pages}{085001}.
\newblock \DOIprefix\doi{10.1103/PhysRevD.85.085001}.
\bibitem[{Bender et~al.(2013)Bender, Branchina and Messina}]{bender2013}
\bibinfo{author}{Bender, C.M.}, \bibinfo{author}{Branchina, V.},
  \bibinfo{author}{Messina, E.}, \bibinfo{year}{2013}.
\newblock \bibinfo{title}{Critical behavior of the
  $\mathcal{P}\mathcal{T}$-symmetric $i{\phi}^{3}$ quantum field theory}.
\newblock \bibinfo{journal}{Phys. Rev. D} \bibinfo{volume}{87},
  \bibinfo{pages}{085029}.
\newblock \DOIprefix\doi{10.1103/PhysRevD.87.085029}.
\bibitem[{Bender et~al.(2005)Bender, Brandt, Chen and Wang}]{bender2005}
\bibinfo{author}{Bender, C.M.}, \bibinfo{author}{Brandt, S.F.},
  \bibinfo{author}{Chen, J.H.}, \bibinfo{author}{Wang, Q.},
  \bibinfo{year}{2005}.
\newblock \bibinfo{title}{The $\mathcal{C}$ operator in
  $\mathcal{P}\mathcal{T}$-symmetric quantum field theory transforms as a
  lorentz scalar}.
\newblock \bibinfo{journal}{Phys. Rev. D} \bibinfo{volume}{71},
  \bibinfo{pages}{065010}.
\newblock \DOIprefix\doi{10.1103/PhysRevD.71.065010}.
\bibitem[{Bender et~al.(2002b)Bender, Brody and Jones}]{bender2002}
\bibinfo{author}{Bender, C.M.}, \bibinfo{author}{Brody, D.C.},
  \bibinfo{author}{Jones, H.F.}, \bibinfo{year}{2002}b.
\newblock \bibinfo{title}{Complex extension of quantum mechanics}.
\newblock \bibinfo{journal}{Phys. Rev. Lett.} \bibinfo{volume}{89},
  \bibinfo{pages}{270401}.
\newblock \DOIprefix\doi{10.1103/PhysRevLett.89.270401}.
\bibitem[{Bender et~al.(2004a)Bender, Brody and Jones}]{bender2004b}
\bibinfo{author}{Bender, C.M.}, \bibinfo{author}{Brody, D.C.},
  \bibinfo{author}{Jones, H.F.}, \bibinfo{year}{2004}a.
\newblock \bibinfo{title}{Extension of $\mathcal{PT}$-symmetric quantum
  mechanics to quantum field theory with cubic interaction}.
\newblock \bibinfo{journal}{Phys. Rev. D} \bibinfo{volume}{70},
  \bibinfo{pages}{025001}.
\newblock \DOIprefix\doi{10.1103/PhysRevD.70.025001}.
\bibitem[{Bender et~al.(2004b)Bender, Brody and Jones}]{bender2004a}
\bibinfo{author}{Bender, C.M.}, \bibinfo{author}{Brody, D.C.},
  \bibinfo{author}{Jones, H.F.}, \bibinfo{year}{2004}b.
\newblock \bibinfo{title}{Scalar quantum field theory with a complex cubic
  interaction}.
\newblock \bibinfo{journal}{Phys. Rev. Lett.} \bibinfo{volume}{93},
  \bibinfo{pages}{251601}.
\newblock \DOIprefix\doi{10.1103/PhysRevLett.93.251601}.
\bibitem[{Bender et~al.(2019)Bender, Dorey, Dunning et~al.}]{bender-book}
\bibinfo{author}{Bender, C.M.}, \bibinfo{author}{Dorey, P.E.},
  \bibinfo{author}{Dunning, C.}, et~al., \bibinfo{year}{2019}.
\newblock \bibinfo{title}{PT Symmetry: In Quantum and Classical Physics}.
\newblock \bibinfo{publisher}{WORLD SCIENTIFIC (EUROPE)}.
\newblock \DOIprefix\doi{10.1142/q0178}.
\bibitem[{Bender and Kuzhel(2012)}]{bender2012}
\bibinfo{author}{Bender, C.M.}, \bibinfo{author}{Kuzhel, S.},
  \bibinfo{year}{2012}.
\newblock \bibinfo{title}{Unbounded -symmetries and their nonuniqueness}.
\newblock \bibinfo{journal}{J. Phys. A: Math. Gen.} \bibinfo{volume}{45},
  \bibinfo{pages}{444005}.
\newblock \DOIprefix\doi{10.1088/1751-8113/45/44/444005}.
\bibitem[{Bender et~al.(2003)Bender, Meisinger and Wang}]{bender2003}
\bibinfo{author}{Bender, C.M.}, \bibinfo{author}{Meisinger, P.N.},
  \bibinfo{author}{Wang, Q.}, \bibinfo{year}{2003}.
\newblock \bibinfo{title}{Calculation of the hidden symmetry operator in
  $\mathcal{PT}$-symmetric quantum mechanics}.
\newblock \bibinfo{journal}{J. Phys. A: Math. Gen.} \bibinfo{volume}{36},
  \bibinfo{pages}{1973}.
\newblock \URLprefix \url{https://dx.doi.org/10.1088/0305-4470/36/7/312},
  \DOIprefix\doi{10.1088/0305-4470/36/7/312}.
\bibitem[{Degrassi et~al.(2012)Degrassi, Di~Vita, Elias-Miro et~al.}]{higgs}
\bibinfo{author}{Degrassi, G.}, \bibinfo{author}{Di~Vita, S.},
  \bibinfo{author}{Elias-Miro, J.}, et~al., \bibinfo{year}{2012}.
\newblock \bibinfo{title}{Higgs mass and vacuum stability in the standard model
  at nnlo}.
\newblock \bibinfo{journal}{JHEP} \bibinfo{volume}{08}.
\newblock \DOIprefix\doi{10.1007/JHEP08(2012)098}.
\bibitem[{Dorey et~al.(2001)Dorey, Dunning and Tateo}]{dorey2001}
\bibinfo{author}{Dorey, P.}, \bibinfo{author}{Dunning, C.},
  \bibinfo{author}{Tateo, R.}, \bibinfo{year}{2001}.
\newblock \bibinfo{title}{Spectral equivalences, bethe ansatz equations, and
  reality properties in $\mathcal{PT}$-symmetric quantum mechanics}.
\newblock \bibinfo{journal}{J. Phys. A: Math. Gen.} \bibinfo{volume}{34},
  \bibinfo{pages}{5679}.
\newblock \DOIprefix\doi{10.1088/0305-4470/34/28/305}.
\bibitem[{Dwivedi and Mandal(2021)}]{aditya2021}
\bibinfo{author}{Dwivedi, A.}, \bibinfo{author}{Mandal, B.P.},
  \bibinfo{year}{2021}.
\newblock \bibinfo{title}{Higher loop $\beta$ function for non-hermitian pt
  symmetric $\iota g\phi^3$ theory}.
\newblock \bibinfo{journal}{Ann. Phys.} \bibinfo{volume}{425},
  \bibinfo{pages}{168382}.
\newblock \DOIprefix\doi{https://doi.org/10.1016/j.aop.2020.168382}.
\bibitem[{Fisher(1978)}]{yanglee}
\bibinfo{author}{Fisher, M.E.}, \bibinfo{year}{1978}.
\newblock \bibinfo{title}{Yang-lee edge singularity and
  ${\ensuremath{\phi}}^{3}$ field theory}.
\newblock \bibinfo{journal}{Phys. Rev. Lett.} \bibinfo{volume}{40},
  \bibinfo{pages}{1610--1613}.
\newblock \URLprefix
  \url{https://link.aps.org/doi/10.1103/PhysRevLett.40.1610},
  \DOIprefix\doi{10.1103/PhysRevLett.40.1610}.
\bibitem[{Fujikawa(1980)}]{fujikawa}
\bibinfo{author}{Fujikawa, K.}, \bibinfo{year}{1980}.
\newblock \bibinfo{title}{Comment on chiral and conformal anomalies}.
\newblock \bibinfo{journal}{Phys. Rev. Lett.} \bibinfo{volume}{44},
  \bibinfo{pages}{1733--1736}.
\newblock \DOIprefix\doi{10.1103/PhysRevLett.44.1733}.
\bibitem[{Jones and Rivers(2007)}]{jones2007}
\bibinfo{author}{Jones, H.F.}, \bibinfo{author}{Rivers, R.J.},
  \bibinfo{year}{2007}.
\newblock \bibinfo{title}{Disappearing $q$ operator}.
\newblock \bibinfo{journal}{Phys. Rev. D} \bibinfo{volume}{75},
  \bibinfo{pages}{025023}.
\newblock \DOIprefix\doi{10.1103/PhysRevD.75.025023}.
\bibitem[{Mostafazadeh(2002a)}]{ali2002-2}
\bibinfo{author}{Mostafazadeh, A.}, \bibinfo{year}{2002}a.
\newblock \bibinfo{title}{Pseudo-hermiticity versus pt-symmetry. ii. a complete
  characterization of non-hermitian hamiltonians with a real spectrum}.
\newblock \bibinfo{journal}{J. Math. Phys.} \bibinfo{volume}{43},
  \bibinfo{pages}{2814--2816}.
\newblock \DOIprefix\doi{10.1063/1.1461427}.
\bibitem[{Mostafazadeh(2002b)}]{ali2002-3}
\bibinfo{author}{Mostafazadeh, A.}, \bibinfo{year}{2002}b.
\newblock \bibinfo{title}{Pseudo-hermiticity versus pt-symmetry iii:
  Equivalence of pseudo-hermiticity and the presence of antilinear symmetries}.
\newblock \bibinfo{journal}{J. Math. Phys.} \bibinfo{volume}{43},
  \bibinfo{pages}{3944--3951}.
\newblock \DOIprefix\doi{10.1063/1.1489072}.
\bibitem[{Mostafazadeh(2002c)}]{ali2002-1}
\bibinfo{author}{Mostafazadeh, A.}, \bibinfo{year}{2002}c.
\newblock \bibinfo{title}{Pseudo-hermiticity versus pt symmetry: The necessary
  condition for the reality of the spectrum of a non-hermitian hamiltonian}.
\newblock \bibinfo{journal}{J. Math. Phys.} \bibinfo{volume}{43},
  \bibinfo{pages}{205--214}.
\newblock \DOIprefix\doi{10.1063/1.1418246}.
\bibitem[{Mostafazadeh(2003)}]{ali2003}
\bibinfo{author}{Mostafazadeh, A.}, \bibinfo{year}{2003}.
\newblock \bibinfo{title}{Exact pt-symmetry is equivalent to hermiticity}.
\newblock \bibinfo{journal}{J. Phys. A: Math. Gen.} \bibinfo{volume}{36},
  \bibinfo{pages}{7081}.
\newblock \DOIprefix\doi{10.1088/0305-4470/36/25/312}.
\bibitem[{Novikov(2019)}]{novikov2019}
\bibinfo{author}{Novikov, O.O.}, \bibinfo{year}{2019}.
\newblock \bibinfo{title}{Scattering in pseudo-hermitian quantum field theory
  and causality violation}.
\newblock \bibinfo{journal}{Phys. Rev. D} \bibinfo{volume}{99},
  \bibinfo{pages}{065008}.
\newblock \DOIprefix\doi{10.1103/PhysRevD.99.065008}.
\bibitem[{Shalaby(2017)}]{shalaby2017}
\bibinfo{author}{Shalaby, A.M.}, \bibinfo{year}{2017}.
\newblock \bibinfo{title}{Vacuum structure and
  $\mathcal{P}\mathcal{T}$-symmetry breaking of the non-hermetian
  ($i{\phi}^{3}$) theory}.
\newblock \bibinfo{journal}{Phys. Rev. D} \bibinfo{volume}{96},
  \bibinfo{pages}{025015}.
\newblock \DOIprefix\doi{10.1103/PhysRevD.96.025015}.
\bibitem[{Shalaby(2019)}]{shalaby2019}
\bibinfo{author}{Shalaby, A.M.}, \bibinfo{year}{2019}.
\newblock \bibinfo{title}{Effective action study of the
  $\mathcal{PT}$-symmetric $i(\phi^3)_{6-\epsilon}$ theory and the yang–lee
  edge singularity}.
\newblock \bibinfo{journal}{Int. J. Mod. Phys. A} \bibinfo{volume}{34},
  \bibinfo{pages}{1950090}.
\newblock \DOIprefix\doi{10.1142/S0217751X19500908}.
\bibitem[{Shalaby(2020)}]{shalaby2020}
\bibinfo{author}{Shalaby, A.M.}, \bibinfo{year}{2020}.
\newblock \bibinfo{title}{Extrapolating the precision of the hypergeometric
  resummation to strong couplings with application to the
  $\mathcal{PT}$-symmetric $i\phi^3$ field theory}.
\newblock \bibinfo{journal}{Int. J. Mod. Phys. A} \bibinfo{volume}{35},
  \bibinfo{pages}{2050041}.
\newblock \DOIprefix\doi{10.1142/S0217751X20500414}.
\bibitem[{Weinberg(1995)}]{weinberg}
\bibinfo{author}{Weinberg, S.}, \bibinfo{year}{1995}.
\newblock \bibinfo{title}{The Quantum Theory of Fields Vol. 1: Foundations.}
\newblock \bibinfo{publisher}{Cambridge University Press}.
\bibitem[{Xian et~al.(2023)Xian, Fernández, Chen et~al.}]{david2023}
\bibinfo{author}{Xian, Z.Y.}, \bibinfo{author}{Fernández, D.R.},
  \bibinfo{author}{Chen, Z.}, et~al., \bibinfo{year}{2023}.
\newblock \bibinfo{title}{Electric conductivity in non-hermitian holography}.
\newblock \href{http://arxiv.org/abs/2304.11183}{{\tt arXiv:2304.11183}}.

\end{thebibliography}

\end{document}